\documentclass[12pt]{article}
\usepackage{amsmath,amsfonts,amssymb,amsthm}
\usepackage{enumerate, multirow}
\usepackage{bm, upgreek}
\usepackage{natbib} 
\usepackage{url,hyperref} 
\usepackage{caption, color, graphicx}
\usepackage{algorithm,algorithmic}  
\usepackage{longtable}
\usepackage{booktabs}

\theoremstyle{plain}

\newtheorem{proposition}{Proposition}

\theoremstyle{definition}

\newcommand{\X}{{\bf X}}
\newcommand{\x}{{\bm{x}}}
\newcommand{\z}{{\bm{z}}}

\newcommand{\Z}{{\bf Z}}

\newcommand{\B}{{\bf B}}
\newcommand{\D}{{\bf D}}

\newcommand{\R}{\mathbb{R}}

\newcommand{\W}{{\bf W}}
\newcommand{\U}{{\bf U}}
\newcommand{\bu}{{\bf u}}
\newcommand{\bv}{{\bf v}}
\newcommand{\V}{{\bf V}}

\newcommand{\bbeta}{\bm{\beta}}
\newcommand{\bEta}{\bm{\eta}}

\newcommand{\Sig}{\bm{\Sigma}}

\newcommand{\bLambda}{\bm{\Lambda}}

\newcommand{\I}{{\bf I}}

\DeclareMathOperator*{\argmin}{arg\,min}
\DeclareMathOperator{\E}{E}

\newcommand{\blind}{0}

\begin{document}

\def\spacingset#1{\renewcommand{\baselinestretch}%
{#1}\small\normalsize} \spacingset{1}


\if0\blind
{
  \title{\bf Adaptive Influence Diagnostics in High-Dimensional Regression}
  \author{Abdul-Nasah, Soale  \thanks{
    Corresponding Author: \textit{abdul-nasah.soale@case.edu}}\hspace{.2cm} \\
  Department of Mathematics, Applied Mathematics, and Statistics, \\
  Case Western Reserve University, Cleveland, OH, USA\\
  Adewale Lukman \\
  Department of Mathematics and Statistics, \\ University of North Dakota, Grand Forks, ND, USA
 }
    \maketitle
} \fi

\if1\blind
{
  \bigskip
  \bigskip
  \bigskip
  \begin{center}
    {\LARGE\bf Title}
\end{center}
  \medskip
} \fi

\bigskip
\begin{abstract}
An adaptive Cook’s distance (ACD) for diagnosing influential observations in high-dimensional single-index models with multicollinearity and outlier contamination is proposed. ACD is a model-free technique built on sparse local linear gradients to temper leverage effects. In simulations spanning low- and high-dimensional design settings with strong correlation, ACD based on LASSO (ACD-LASSO) and SCAD (ACD-SCAD) penalties reduced masking and swamping relative to classical Cook's distance and local influence as well as the DF-Model and Case-Weight adjusted solution for LASSO. Trimming points flagged by ACD stabilizes variable selection while preserving core signals. Applications to two datasets--the 1960 US cities pollution study and a high-dimensional riboflavin genomics experiment show consistent gains in selection stability and interpretability. 
\end{abstract}

\noindent%
\vfill

\newpage
\spacingset{2} 

\section{Introduction \label{sec:1}}
Effective regression analysis requires rigorous diagnostics to assess the adequacy of the assumed model and the stability of estimates, which is difficult to achieve when \emph{influential observations} are present in the data. An observation is considered influential if its inclusion or omission significantly alters the regression estimates \citep{cook1977detection}. Thus, a small fraction of such atypical data points can dramatically result in misleading inference, variable selection, and prediction. Diagnosing influential observations has received a lot of attention in the literature from classical linear and single-index models \citep{cook1977detection, Belsley1980, prendergast2008trimming} to Fr\'echet regression \citep{soale2025detecting}. 

Existing influence diagnostics can be broadly categorized into two: the \emph{case-deletion} approach and the \emph{perturbation} method. The case-deletion method involves quantifying the effect of omitting one or more observations on the regression estimates. This technique dates back to Cook's distance, DFBETAS, and DFFITS \citep{cook1977detection, cook1980c, chatterjee1986}, which are based on \emph{single-case deletion}. The single-case deletion is shown to suffer from masking (failure to detect a true influential point) and swamping (falsing flagging a normal point as influential) due to adjacent outliers, especially, when the proportion of influential observations in the data is substantial \citep{Belsley1980}. Thus, the \emph{multiple-case deletion} was proposed, see \cite{Belsley1980,pena2005,pena1999fast,nurunnabi2014procedures,roberts2015adaptive}, to name a few. On the other hand, the perturbation method involves adding minor perturbations in the model, usually in the predictors or the response, to assess the effect of the perturbation on the regression estimates. Earlier works in this directions can be traced to \citep{hodges1972data, Belsley1980, polasek1984regression, cook1986assessment}. 

Modern data analysis typically involves sparse high-dimensional regression, which has been shown to suffer when atypical observations are present in the data. For example, sparsity–inducing methods such as the lasso \citep{tibshirani1996regression} are highly sensitive to single observations, which can alter both coefficient magnitudes and model selection \citep{kim2015case, rajaratnam2019influence}. Nonconvex penalties such as SCAD \citep{fan2001variable} and MCP \citep{zhang2010} reduce shrinkage bias but remain vulnerable to leverage points and heavy–tailed errors. This has prompted the development of influence diagnostics measures for high-dimensional regression. For example, \cite{wang2019multiple} extended the adaptive automatic multiple-case deletion of \cite{roberts2015adaptive} to sparse high-dimensional linear models. Similarly, \cite{lu2022sparse} proposed the sparse local influence method for high-dimensional regression. Recently, \cite{rajaratnam2019influence} and \cite{jiao2025assessment} proposed the DF-model and case-weight adjusted solution for LASSO regression, respectively. 

Yet, regardless of the paradigm adopted for influence detection, most existing methods share two limitations in high-dimensional settings. First, they are handicapped by severe collinearity in the predictors, which exacerbates the masking and swamping problem. This is particularly pronounced in the LASSO-based techniques such as the DF-model and case-weight methods as they inherit the grouping effect problem of LASSO. Another common limitation is that they are susceptible to model misspecification as most assume a linear model or a known link. As noted by \cite{cook1986assessment}, the local influence methods rely on a ``well-behaved" likelihood, which is not always guaranteed. 

To address these challenges, we propose an adaptive Cook's distance method which utilizes sparse local linear gradients. Our framework for influence diagnostics is developed for general single index models. Thus, we do not require specification of the response link. The adaptive Cook’s distance exploits the connection between case-deletion and local linear approximation to stabilize influence assessment under collinearity and contamination. Kernel weights are employed to dampen the effect of high-leverage points while flexible sparsity-inducing penalties are incorporated to allow for variable selection in the local estimation. The index direction is estimated as the leading principal component of the sparse local gradients, which is then used to compute the adaptive Cook's distance measure. 

The remainder of this article is organized as follows. Section~\ref{sec:2} proposes the model and index estimation via local linear regression. In Section~\ref{sec:3}, we establish the connection between local linear regression and the case-deletion technique, and proceed to introduce the adaptive Cook's distance. Section~\ref{sec:4} reports extensive simulation studies under various contamination and collinearity scenarios, followed by two real–data applications in Section~\ref{sec:5}. Concluding remarks are given in Section~\ref{sec:6}, with technical details and proofs relegated to the Appendix.

\subsection{A Motivating Example}
\begin{figure}[htb!]
    \centering
    \includegraphics[width=0.49\linewidth, height=2.7in]{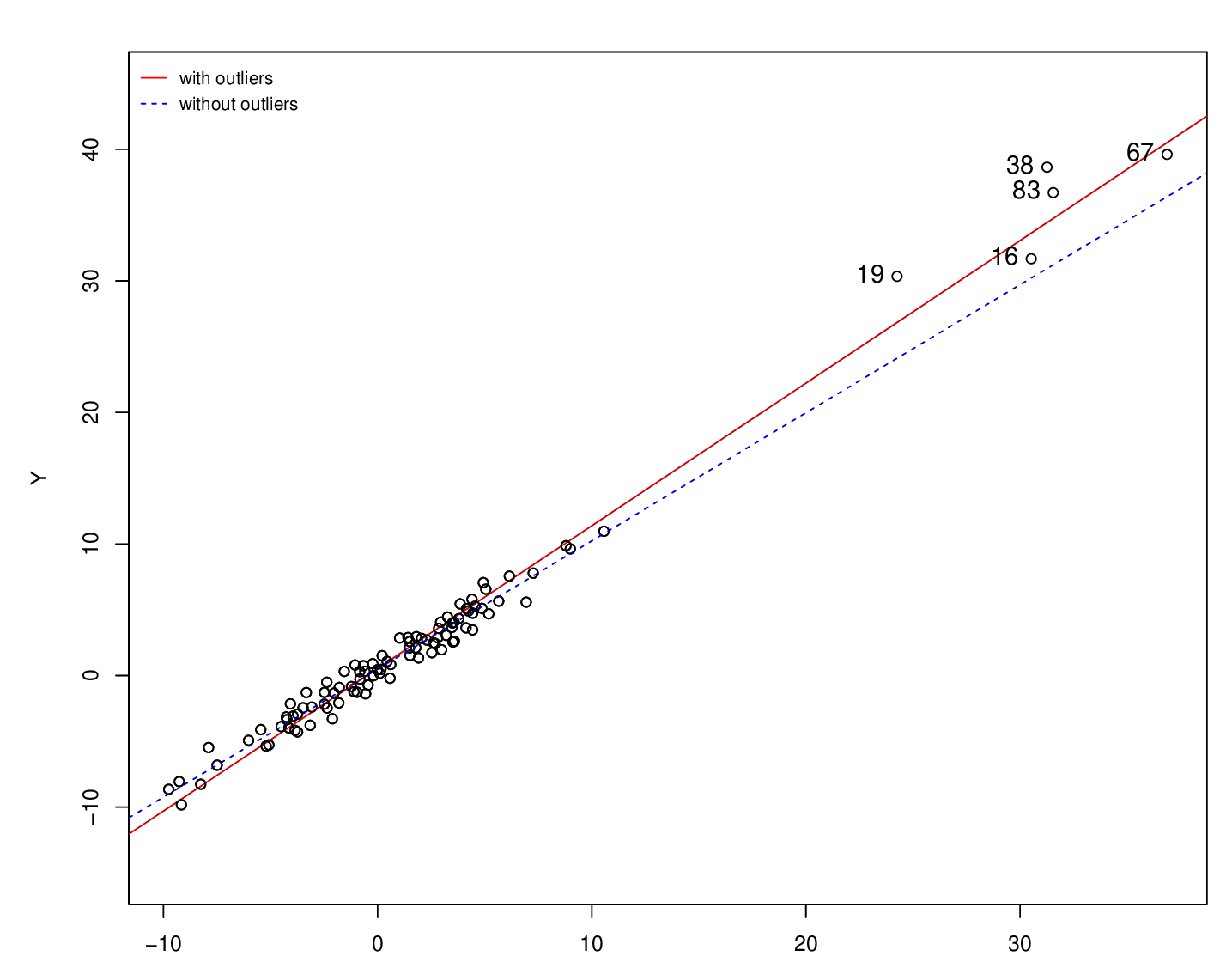}
     \includegraphics[width=0.49\linewidth, height=2.7in]{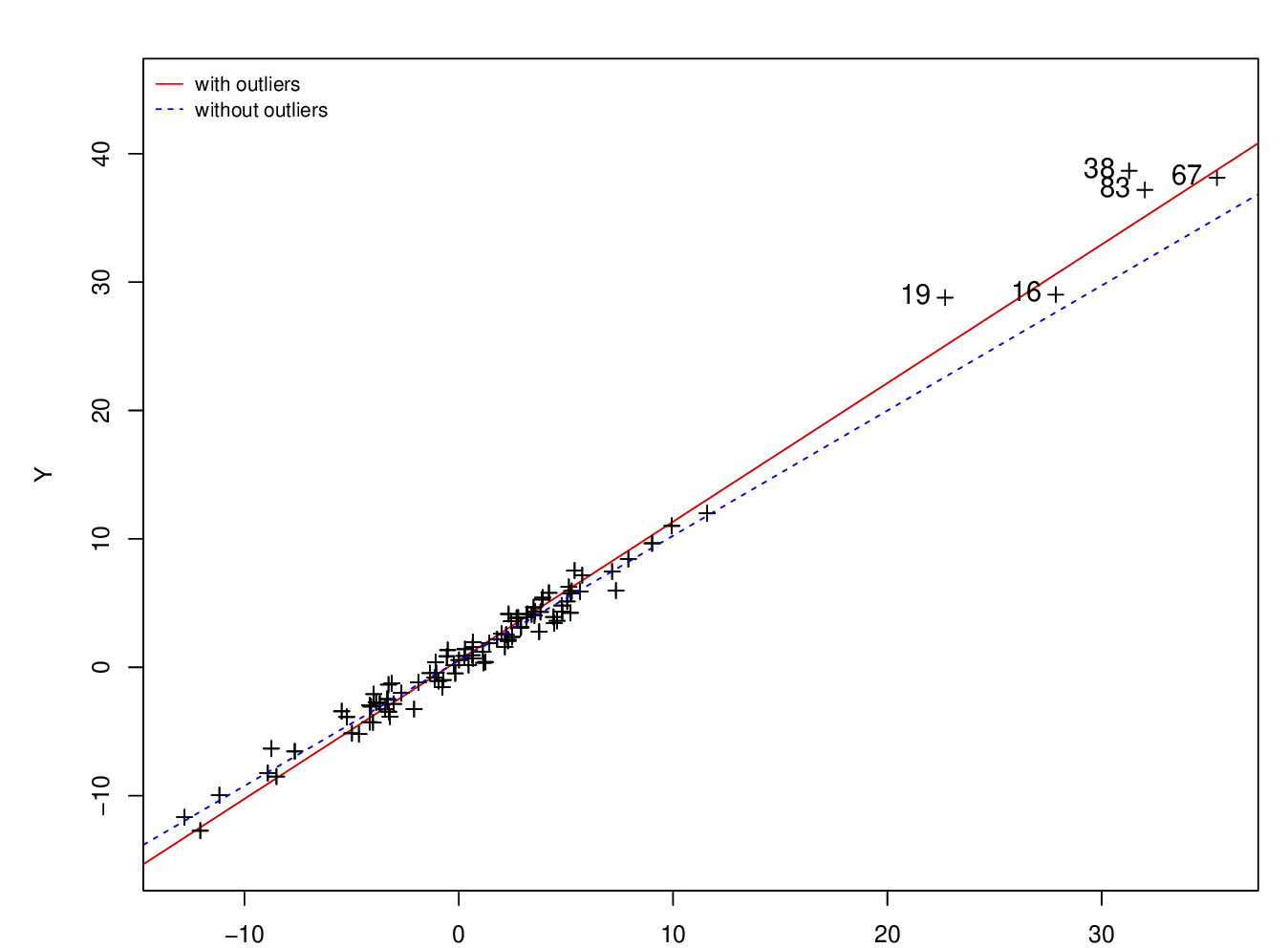}
    \vspace{-0.05in}
    \caption{Scatter plot for design with independent correlation (left) and autoregressive correlation (right), with labels denoting the outliers. Blue line indicates the best fit line in the uncontaminated sample while the red line indicates the shift due to the outliers}
    \label{fig:toy_ex}
\end{figure}

We illustrate the adaptive Cook’s distance using a classical linear regression model contaminated by influential observations under varying correlation structures in the design matrix. Specifically, we set $(n,p) = (100,10)$ and take the regression coefficient vector as $\beta = (3, 1.5, 0, 0, 2, 0, \ldots, 0)^\top \in \mathbb{R}^p$. Of the 100 observations, 95 are generated according to 
\[
\x_i \sim N_p(\mathbf{0}, \Sig), 
\qquad 
y_i = 0.5 + \x_i^\top \beta + 0.8 \varepsilon_{1i}, 
\quad \varepsilon_{1i} \sim N(0,1),
\]
representing the uncontaminated sample. The remaining 5 observations are generated from a heavy-tailed distribution with shifted mean, 
\[
\x_i \sim t_{10}(\mathbf{5}, \Sig), 
\qquad 
y_i = 0.5 + \x_i^\top \beta + \varepsilon_{2i}, 
\quad \varepsilon_{2i} \sim \chi^2(5),
\]
to mimic influential cases. Two covariance structures for $\Sig$ are considered: 
\begin{enumerate}[(a)]
    \item independence, $\Sig = \I_p$, 
    \item autoregressive correlation, $\Sig_{kl} = 0.5^{|k-l|}, \; k,l=1,\ldots,p$.
\end{enumerate} 
A scatter plot illustrating the uncontaminated and contaminated samples is displayed in Figure~\ref{fig:toy_ex}.

\begin{figure}[htb!]
    \centering
    \includegraphics[width=\linewidth,]{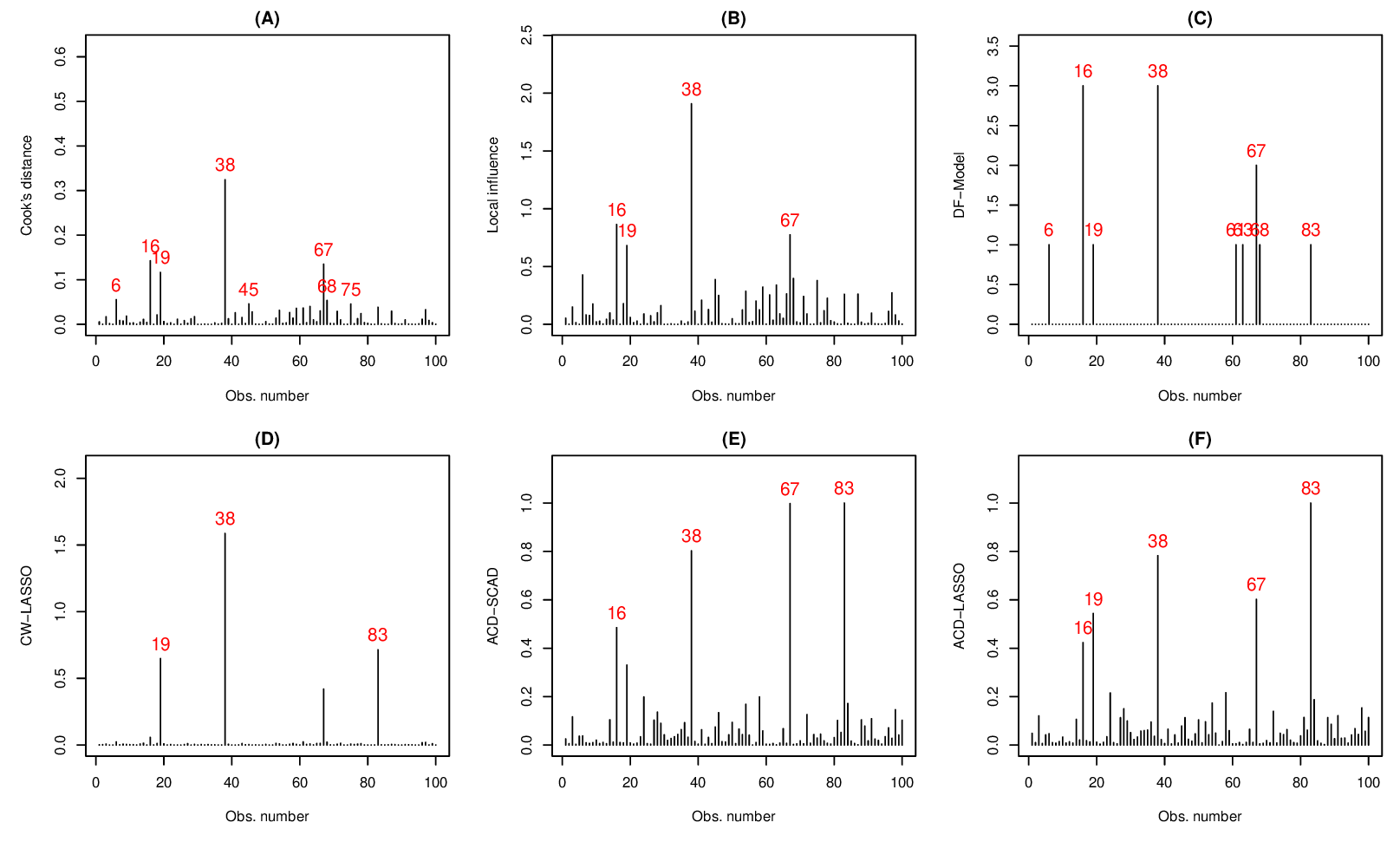}
    \caption{Influence measures under independent design}
    \label{fig:toy_ind}
\end{figure}

\begin{figure}[htb]
    \centering
    \includegraphics[width=\linewidth,]{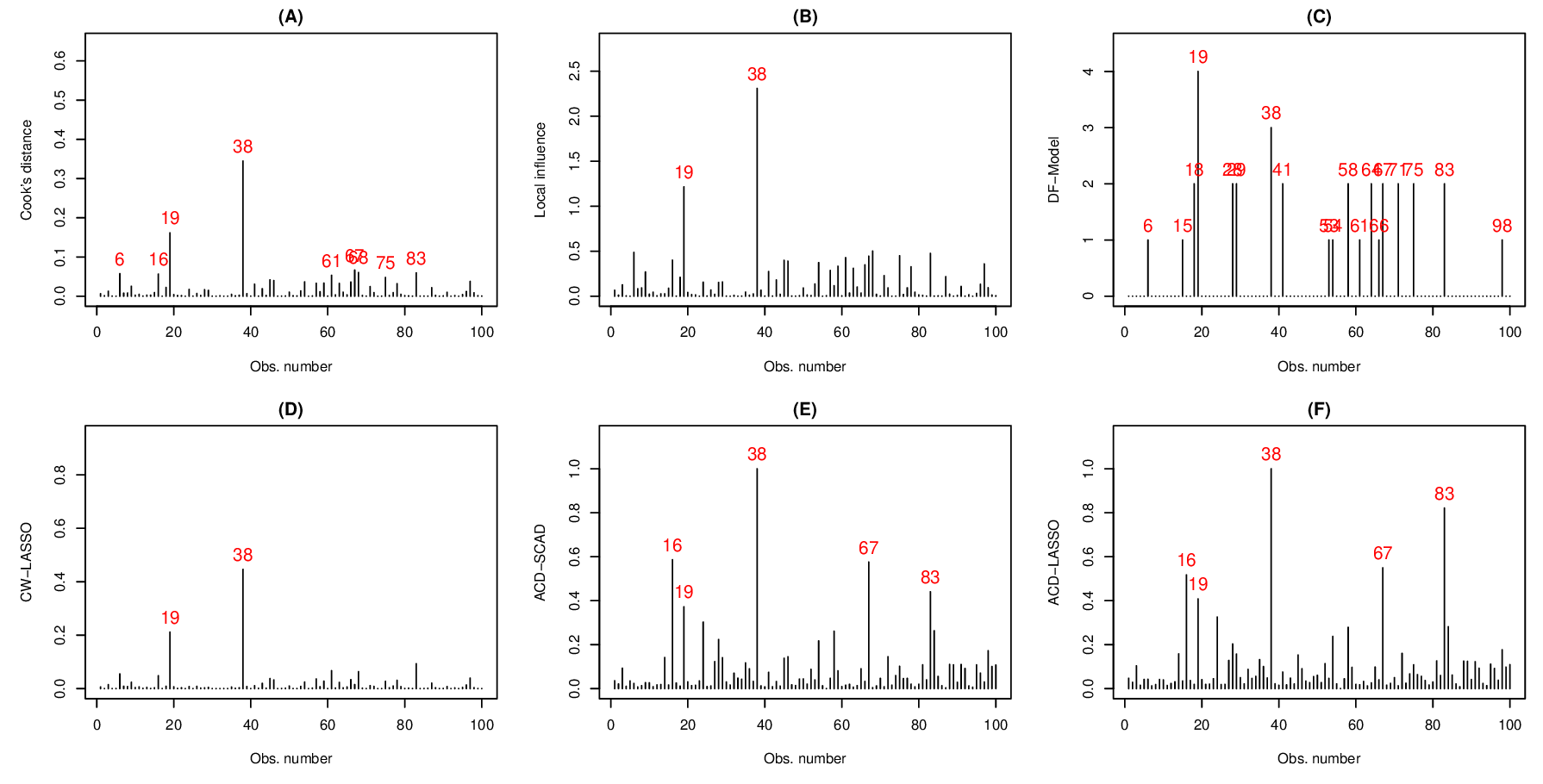}
   \caption{Influence measures under correlated design}
    \label{fig:toy_cor}
\end{figure}

Although the regression coefficients and the set of outliers are the same across both designs, the relative performance of the influence diagnostics differ markedly (Figures~\ref{fig:toy_ind}--\ref{fig:toy_cor}). Under the independent design, Cook’s distance \citep{cook1977detection} and local influence \citep{cook1986assessment} correctly identified most of the outliers, with the exception of observation 83. The DF-model \citep{rajaratnam2019influence} captured all outliers but suffers from swamping. \cite{jiao2025assessment}'s case-weight adjusted solution for LASSO could not capture observation 16. In contrast, the adaptive Cook’s distance with LASSO penalty accurately detected the full set of outliers without false positives and all but observation 19 with the SCAD penalty. 

When the correlated design is introduced, Cook’s distance and the DF-model become less reliable, struggling more with swamping. However, the local influence and case-weight for LASSO responded in the opposite direction, suffering more from masking. In comparison, the adaptive Cook’s distance remained stable and accurate, consistently detecting only the true outliers. Increasing the correlation to $0.7$ and $0.9$ further amplified the shortcomings of the existing methods. In all cases, the adaptive Cook’s distance retained its robustness, providing reliable identification of influential cases even in the presence of multicollinearity. In the sections that follow, we will describe the development of the adaptive Cook's distance. 

\section{Model\label{sec:2}}
For a scalar response $Y\in \R$ and a $p$-dimensional predictor $\X\in\R^p$, consider the single-index model:
\begin{align}
    Y = f(\bbeta^\top\X) + \epsilon,
    \label{mode11}
\end{align}
where the function $f(.)$ is unknown, the parameter $\bbeta\in\R^p$ with $\lVert \bbeta \rVert = 1$, the noise $\epsilon$ is such that $\E(\epsilon|\X)=0$, and $\lVert.\rVert$ denotes the $\ell_2$-norm. In generalized linear models, we assume $f(.)$ is known, which can easily lead to model misspecification. Thus, here we avoid such restrictions. 

By (\ref{mode11}), we also have that $f(\bbeta^\top\X) = \E(Y|\X)$, which can be approximated using local linear regression. Assuming $f$ is differentiable at $\X = \x_0$, the conditional mean at this point can be approximated using the first-order Taylor expansion: 
\begin{align}
    \E(Y|\X=\x_0) \approx a + \B^\top (\X-\x_0),
    \label{approx}
\end{align}
where $a =  f(\bbeta^\top\x_0) \in \R$ and $\B=\cfrac{\partial f(\bbeta^\top\X)}{\partial \X}\Bigm|_{\X=\x_0} \in \R^p$. 

\begin{proposition}\label{opg}
    Suppose $f$ is differentiable. Then $\bbeta$ is in the space spanned by the leading eigenvector of $\E(\B\B^\top)$.
\end{proposition}
\noindent Proposition \ref{opg} follows directly from Lemma 1 of \cite{xia2002adaptive}. See also \cite{li2018sufficient} for more details on outer-product of gradients for sufficient dimension reduction.  

To obtain the estimates for $(a,\B)^\top$ at $\x_0$, we solve  
\begin{align}
    \argmin_{a,\B}\E[(Y-a-\B^\top(\X-\x_0))^2w(\x_0)],
    \label{loc_linreg}
\end{align}
where $w(\x_0)$ denotes some weighting function. Usually $w(\x_0) = K_h((\X-\x_0)/h)$, where $K_h(.)$ denotes a kernel function with bandwidth $h$.  While there are several options for $K_h(.)$, our kernel of choice is the Gaussian radial basis function given by $K_h(u) = \cfrac{1}{\sqrt{2\pi}}\exp(-u^2/2h)$. This kernel gives lower weights to points that are farther from $\x_0$, thus, dampening the effect of distant high leverage points on the local estimate. Moreover, by choosing an optimal radius, we can obtain results similar to the multiple-case deletion. 

\section{Adaptive Cook's Distance\label{sec:3}}
\subsection{Single-case Deletion Revisited}
Consider the classical linear model $Y = \beta_0 + \beta_1X_1 + \ldots + \beta_pX_p + \epsilon$, where $\E(\epsilon|\X)=0$. Given a random sample $(\x_1,y_1),\ldots, (\x_n,y_n)$ of $(\X,Y)$, the ordinary least squares (OLS) estimator is given by 
\begin{align}
  \hat\bbeta = (\X_n^\top\X_n)^{-1}\X_n^\top Y_n,
 \label{ols_est}
\end{align}
where $\X_n = (\bm 1, \x_1,\ldots,\x_n)^\top$, $Y_n=(y_1,\ldots,y_n)^\top$, and $\hat\bbeta = (\hat\beta_0, \hat\beta_1,\ldots,\hat\beta_p)^\top$. It is well-known that $\hat\bbeta$ is susceptible to both outliers and  collinearity in $\X$. Thus, the presence of a few influential observations coupled with mild collinearity can lead to unstable estimates, and consequently, misleading inference \citep{kim2015case, rajaratnam2019influence}. 

To measure the influence of an individual observation, a popular technique is to use the single-case deletion approach to see if there is a significant change in $\hat\bbeta$ or the predicted response $\hat Y_n = \hat\bbeta^\top\X_n$. Suppose $(\X_{-i}, Y_{-i}) = (\X_n, Y_n)\backslash \{(\x_i, y_i)\}$, i.e., the sample without the $i$th observation. Then the corresponding OLS estimator based on $(\X_{-i}, Y_{-i})$ is given by
\begin{align}
    \hat\bbeta^{-i} = (\X_{-i}^\top\X_{-i})^{-1}\X_{-i}^\top Y_{-i} = \hat\bbeta - \cfrac{(\X_n^\top\X_n)^{-1}\x_i}{1-h_{ii}}e_i,
    \label{beta_i}
\end{align}
where $h_{ii} = \x_i^\top(\X_n^\top\X_n)^{-1}\x_i$ is the {\it leverage} and $e_i = y_i -\hat y_i$ is the residual of the $i$th observation. From (\ref{beta_i}), we see that deviation between $\hat\bbeta^{-i}$ and $\hat\bbeta$ can come from high leverage, i.e., $h_{ii} \to 1$, high residual value, ill-conditioning of $\X_n^\top\X_n$, or a combination of these factors. Moreover, as both $\hat\bbeta$ and $\hat\bbeta^{-i}$ depend on $(\X_n^\top\X_n)^{-1}$, severe multicollinearity in $\X_n$ can lead to ill-conditioning of $\X_n^\top\X_n$, making the inverse highly unstable. In such a scenario, a small change in the $\x_i$'s can result in huge changes in  $(\X_n^\top\X_n)^{-1}$, and consequently $\hat\bbeta$. 

Consider the case of the Cook's distance, which for the $i$th observation is given by 
\begin{align}
    D_i = \cfrac{(\hat\bbeta - \hat\bbeta^{-i})^\top\X_n^\top\X_n(\hat\bbeta - \hat\bbeta^{-i})}{p\hat\sigma^2},
    \label{cooksdist}
\end{align}
where $\hat\bbeta^{-i}$ is the OLS estimate without the $i$th observation and $\hat\sigma^2$ is the estimated error variance based on all observations \citep{cook1977detection}. If $\X_n$ exhibits multicollinearity, we expect the performance of this diagnostics measure to degrade. Sadly, the local influence measures and the recent extensions such as the DF-model and case-weight for LASSO are also not immune to this effect of multicollinearity as illustrated in the motivating example.

\subsection{Case-Deletion: A Special Case of Local Linear Regression}

The case-deletion technique is a special case of the local linear regression. To see this, notice that the estimator $\hat\bbeta^{-i}$ given in (\ref{beta_i}) can also be expressed as
\begin{align}
  \hat\bbeta^{-i} = (\X_n^\top \W_i \X_n)^{-1}\X_n^\top \W_i Y_n,
\end{align}
where the weighting function $\W_i=diag(w_{jj})$ is a diagonal matrix of binary weights with $w_{jj} = 0 \text{ if } j = i \text{ and } 1 \text{ if } j \neq i$. Similarly, the multiple case deletion is equivalent to setting the weights to 0 for all the deleted observations and 1 otherwise.

If we generalize the binary weights $\W_i$ to some function $w(\x_i)$, $\hat\bbeta^{-i}$ becomes the solution to the local linear regression in (\ref{loc_linreg}). By adopting the kernel weights, we can improve the case deletion estimator $\hat\bbeta^{-i}$ by choosing the weight function to dampen the effect of distant high leverage points. Also, using kernel weights can help deal with multicollinearity as variables that are globally collinear may not necessarily exhibit the same level of collinearity in the local neighborhood of $\x_i$. That is, $\X_n^\top\W_i\X_n$ is less likely to be ill-conditioned under global collinearity compared to $\X_n^\top\X_n$, although the former is not completely immune.

\subsection{Adaptive Cook's Distance}
Another obvious limitation of the original Cook's distance besides multicollinearity is that it is not directly applicable in very high-dimensional settings, i.e., $p > n$. To address this, the adaptive Cook's distance (ACD) not only exploits the connection between case deletion and local linear regression, but also impose sparsity on $\B$. Thus, we solve the penalized local weighted least squares: 
\begin{align}
   (\hat a_i, \hat \B_i)^\top = \argmin_{a,\B}  \displaystyle\sum_{j=1}^n \big[y_j - a - \B^\top(\x_j-\x_i) \big]^2 w(\x_i)  + P_\lambda(\B), 
    \label{sparse_gradients} 
\end{align}
where $P_\lambda(\B)$ is a sparsity-inducing penalty term and the weight $w(\x_i)$ is given by
\begin{align*}
     w(\x_i) = \cfrac{\exp(-\lVert \x_i-\x_j \rVert/\uptau^2)}{\displaystyle\sum_{j=1}^n \exp(-\lVert \x_i-\x_j \rVert/\uptau^2)}.
\end{align*}
The penalty $P_{\lambda}(\B)$ can be any of the popular sparsity-inducing functions. Here, 
we will only focus on LASSO and SCAD, which are given by
\begin{align}
 \begin{cases}
      P_{\lambda}(\B) = \lambda\displaystyle\sum_{k=1}^p |B_k| \text{ for LASSO}, \\
      P_{\lambda}(B_k) =  \lambda{\displaystyle\int}_{0}^{|B_k|}\min{\big\{}1, (\gamma-t/\lambda)_+/(\gamma-1) {\big\}} dt \text{ for SCAD},
  \end{cases}  
\end{align}
where $(c)_+ = \max(0, c)$ with $\gamma > 2$, typically set around 3.7. LASSO is effective at performing variable selection but tends to give biased estimates when there is strong multicollinearity as it arbitrarily selects one of the variables in the correlated group \citep{zou2005regularization}. On the other hand, SCAD is designed to reduce bias and achieves ``oracle properties" asymptotically \citep{fan2001variable}.  

Let $\hat\bEta_i = (\hat a_i, \hat \B_i)^\top$  and $\bLambda_n = (\hat\bEta_1,\ldots,\hat\bEta_n)^\top \in \R^{n \times (p+1)}$ be the vector of sparse local parameter estimates. By the reduced singular value decomposition (SVD), if $\bLambda_n$ has rank $r$, then $\bLambda_n$ can be decomposed as $\bLambda_n = \U\D\V^\top$,
where $\U = (\bu_1,\ldots,\bu_r) \in \R^{n\times r}$ and $\V^\top= (\bv_1,\ldots,\bv_{p+1}) \in\R^{r\times (p+1)}$ are orthonormal vectors, and $\D=diag(d_1,\ldots,d_r) \in\R^{r\times r}$ with $d_1 \geq d_2, \ldots,\geq d_r > 0$. Thus, the columns of $\V$ are the principal component loadings of $\bLambda_n^\top\bLambda_n$. Following Proposition 1, we retain only $\bv_1$, the eigenvector corresponding to the leading eigenvalue.

By replacing $\hat\bbeta^{-i}$ with $\hat\bEta_i$ and $\bbeta$ with $\bv_1$, we define the adaptive Cook's distance for the $i$th observation as
\begin{align}
    D_i^A = \cfrac{(\bv_1 - \hat\bEta_i)^\top\X_n^\top\X_n(\bv_1 - \hat\bEta_i)}{pd_1^2},
    \label{adapt_cd}
\end{align}
where the $\hat\sigma^2$ in (\ref{cooksdist}) is replaced by the corresponding eigenvalue $d_1^2$ of $\bv_1$. Note that because we are dealing with a squared distance, the lack of identifiability of $\bv_1$ should not affect $D_i^A$.\footnote{$\bv_i$'s are only unique up to a sign flip.} 

Once, we compute $\D^A_n = (D_1^A, \ldots, D_n^A)^\top$, we proceed to scale the distances using the min-max normalization to obtain
\begin{align}
    \tilde{\D}^A_n = \cfrac{\D^A_n - \min(\D^A_n)}{\max(\D^A_n) - \min(\D^A_n)},
\end{align}
with $0 \leq \tilde{D}_i^A\leq 1, \ i=1,\ldots,n$. In the spirit of \cite{rajaratnam2019influence}, we consider observations with $\tilde{D}_i^A > mean(\tilde{\D}^A_n) + 2SD(\tilde{\D}^A_n)$ as influential. However, the investigator may choose a reasonable fixed cutoff, say, 0.5.

The adaptive Cook's distance is summarized in Algorithm \ref{alg_acd}. To ensure fair penalization, all the predictors are standardized to be on the same scale. Also, the algorithm relies on a number of tuning parameters, which need to be addressed in the implementation. First, the kernel weight in Step 2 depends on the parameter $\uptau$. To determine the optimal value, we follow the recommendation of \cite{li2011principal} and compute it as 
\begin{align*}
    \hat\uptau = \cfrac{1}{\binom{n}{2}}\displaystyle\sum_{i < j, \ j=2}^n \lVert \x_{i} -\x_{j}\rVert.
\end{align*}
\noindent Next, the regularization parameter $\lambda$ in Step 4 is determined using the usual cross-validation over a grid of values. This can be done using standard packages such as {\it glmnet} or {\it ncvreg} in $\mathrm{R}$ statistical software.\\

\begin{algorithm}
\caption{Adaptive Cook's Distance}
\label{alg_acd}
\begin{algorithmic}[1]
\STATE Input $(\X_n, Y_n)$
\STATE Marginally scale $\X_n$ as $\Z_n$ and center $Y_n$ as $Y_c$
\FOR{$i = 1$ to $n$}
    \STATE Compute ${\bm w}_i \gets \cfrac{\exp(-\lVert \z_i-\z_j \rVert/\hat\uptau^2)}{\displaystyle\sum_{j=1}^n \exp(-\lVert \z_i-\z_j \rVert/\hat\uptau^2)}, \ j=1,\ldots,n$
    \STATE Set $\widetilde{Y}_n \gets \sqrt{{\bm w}_i} Y_c$ and $\widetilde{\Z}_n \gets \sqrt{{\bm w}_i} (\Z_n-\z_i)$
    \STATE Compute the LASSO/SCAD estimate $\hat\bEta_i = (\hat a_i, \hat \B_i)^\top$ for regressing $\widetilde{Y}_n$ on $\widetilde{\Z}_n$
\ENDFOR
\STATE Set $\bLambda_n \gets (\hat\bEta_1,\ldots,\hat\bEta_n)^\top$ and a compute $(\hat{\bu}_1, \hat{\bv}_1, \hat d_1)^\top \gets svd(\bLambda_n)$ 
\STATE Compute $\tilde \D_n^A$
\end{algorithmic}
\end{algorithm}

\subsection{Extension to Non-monotone Link Functions}
Because ACD is constructed from local (data–adaptive) gradients of the fitted mean, it is not restricted to linear models or monotone links. In particular, unlike local influence, ACD does not require specifying the form of the link $f(\cdot)$ and can be applied when the mean is a general smooth function of the predictors.

\begin{figure}[htb!]
    \centering
    \includegraphics[width=0.9\linewidth]{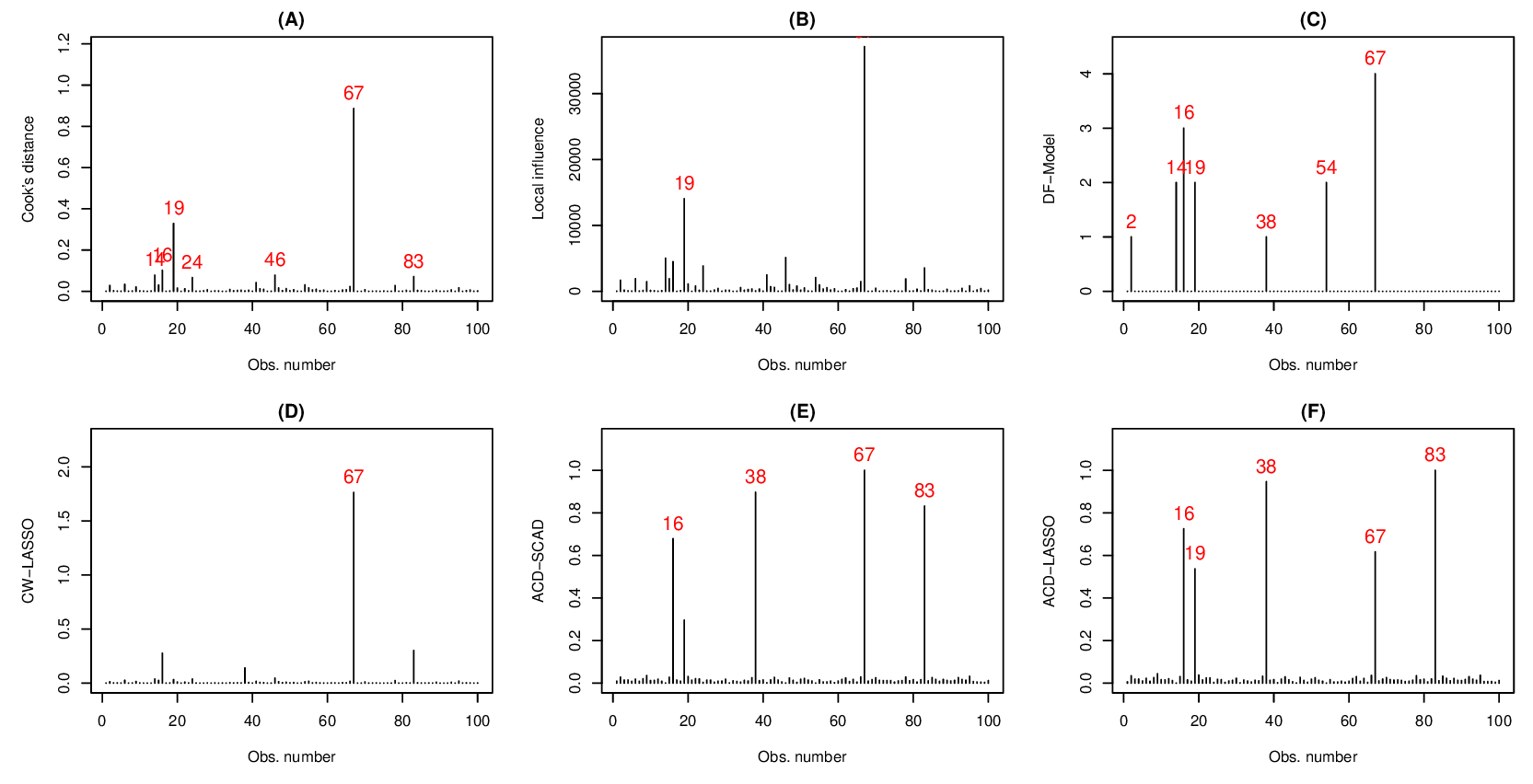}
    \caption{Detecting influential observations under symmetric link}
    \label{fig:nonlinear}
\end{figure}

To illustrate, consider a slight modification of the motivating example:
\[
Y \;=\; \bigl(0.5 + \boldsymbol{\beta}^\top \mathbf{X}\bigr)^2 + \varepsilon,
\]
with the same settings as in the motivating example and $\boldsymbol{\Sigma} = \mathbf{I}_p$. The injected outliers remain $\{16, 19, 38, 67, 83\}$. Owing to the symmetry of the squared link, diagnostics that implicitly assume a linear link—Cook’s distance, local influence, DF–model and CW-LASSO exhibit reduced sensitivity to these outliers (see Fig.~\ref{fig:nonlinear}). By contrast, ACD continues to prioritize the true outliers and avoids spurious flags, highlighting its robustness to model misspecification.

\section{Simulation Study \label{sec:4}}
In this section, we explore the performance of our proposal in diagnosing influential observations in synthetic data compared to existing techniques. Our adaptive Cook's distance based on the LASSO (ACD-LASSO) and SCAD (ACD-SCAD) penalties are considered. Competing methods considered include the original Cook's distance (CKD), local influence measure (LOC-INF), DF-Model, and Case-Weight adjusted method for LASSO (CW-LASSO).  

In the second part of the simulation study, we investigate the effect of trimming the detected influential observations on variable selection. In both cases, we consider low dimensional ($p << n)$ and high dimensional ($p >> n)$ designs under different correlation structures.

\subsection{Detecting Influential Observations}
Here, we consider two models. For each model we include $P_{out}$\% of the observations as outliers. To evaluate performance, let $\mathcal{O} = \{ \text{true outliers} \}$. We report the true positive rate of the detection as 
\[ \text{TPR} = \cfrac{|\{j \in \{1,\ldots,n\}: j \in \mathcal{O}\}|}{|\mathcal{O}|}, \]
where $j$ is an observation flagged as influential and $|.|$ denotes the cardinality of the set. 

{\bf Model I:} Set $(n,p) = (100, 10)$, $\beta = (3,1.5,0,0,2,0,\ldots, 0)^\top \in \R^p$ and $P_{out} = 5\%$. Next, generate 95 observations as $\x_i \sim N_p(\bm 0, \Sig)$ and $y_i = 0.5  + \beta^\top\x_i + 0.8\epsilon_{1i}$, where $\epsilon_1 \sim N(0,1)$; and the remaining 5 observations as $ \x_i \sim t_{10}(\bm 5, \Sig)$ and $y_i = 0.5 + \beta^\top\X_i + \epsilon_{2i}$, where $\epsilon_2 \sim \chi^2(5)$ with 
(a) $\Sig = 0.5^{|k-l|}$,  and (b) $\Sig_{ij} = 1$ if $k=l$ and 0.5 otherwise, where $k,l=1,\ldots,p$.

\begin{figure}[htb!]
    \centering
    \includegraphics[width=0.49\linewidth, height=3in]{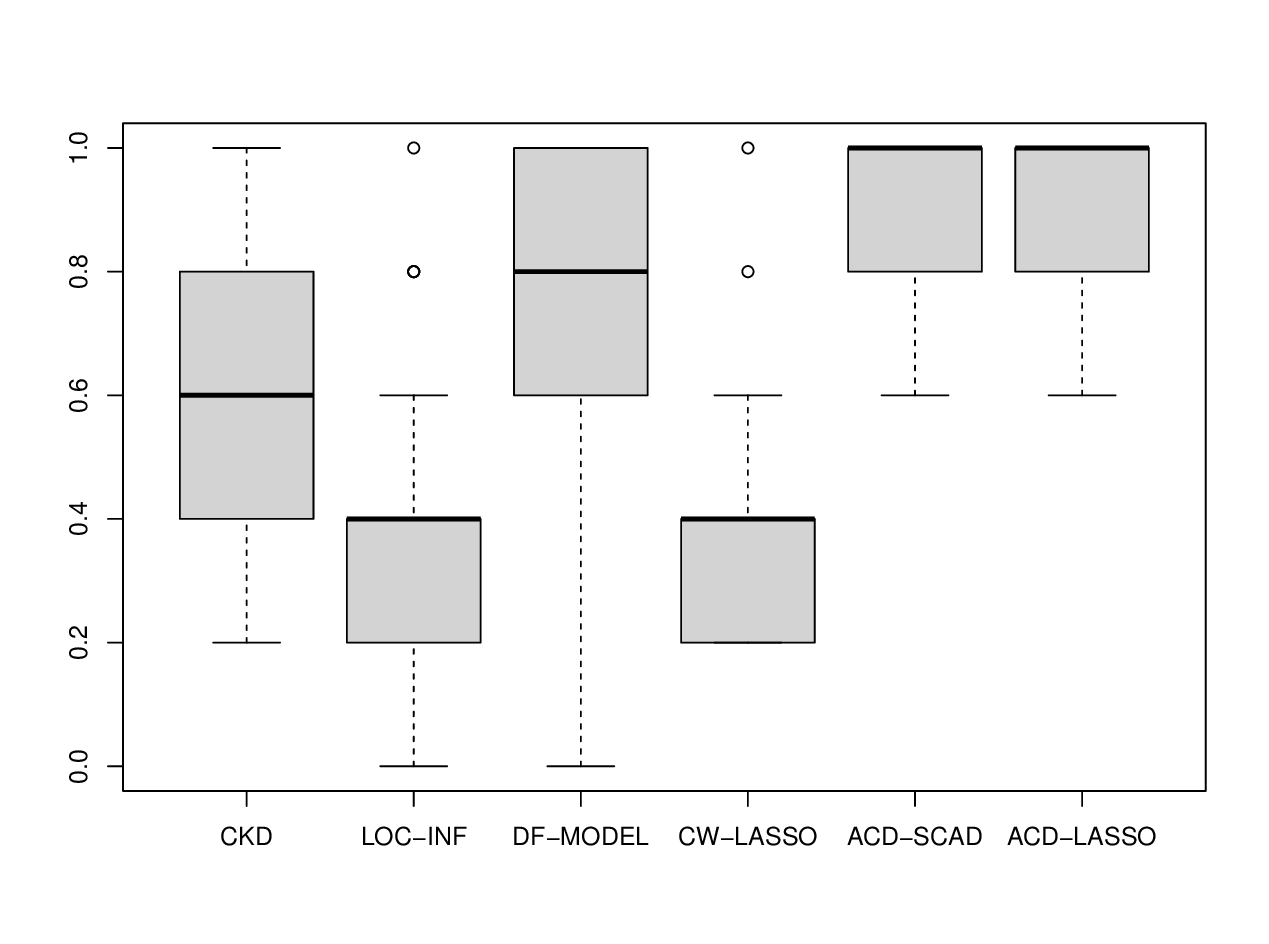}
     \includegraphics[width=0.49\linewidth, height=3in]{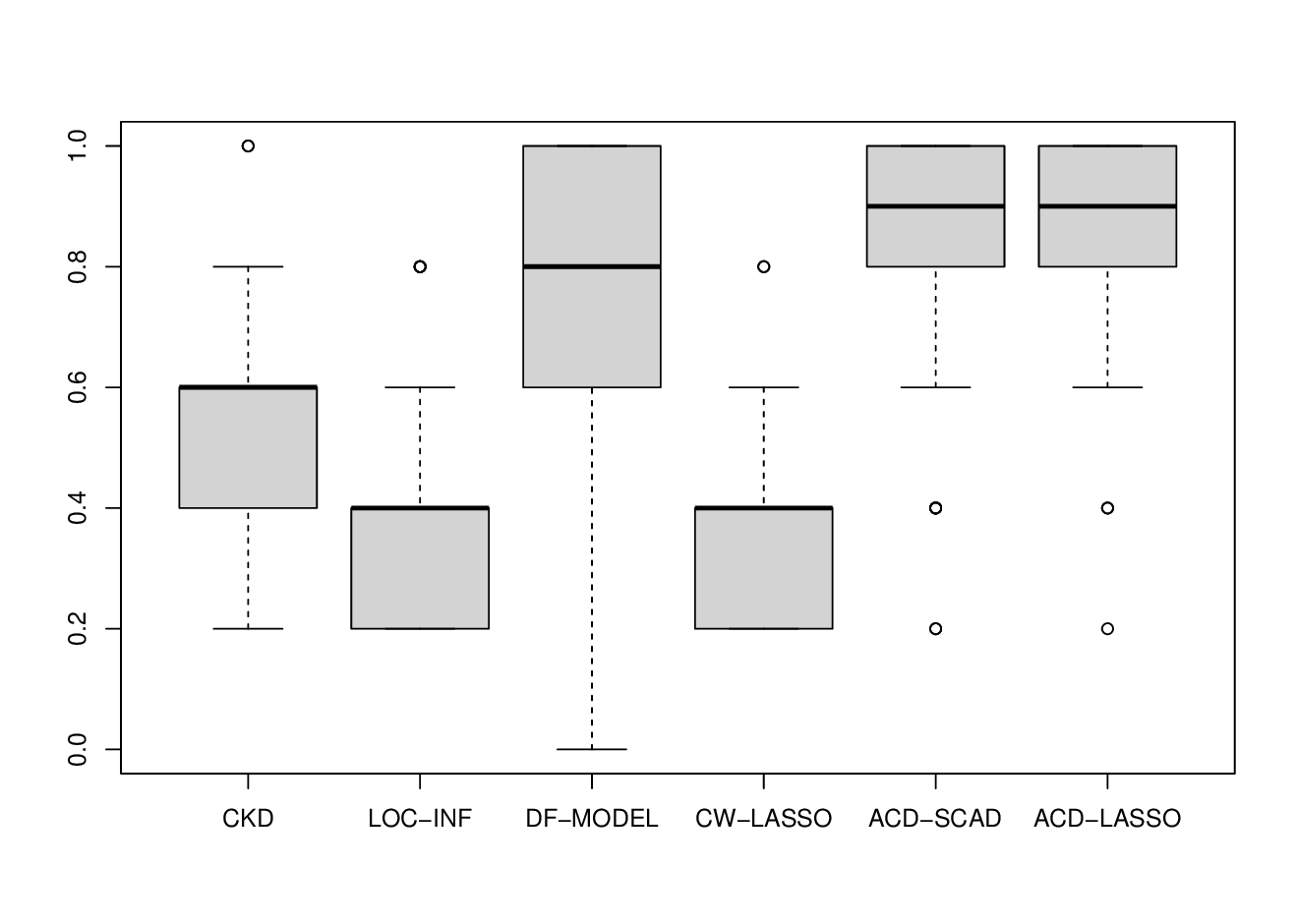}
    \caption{TPR of influence detection in low dimension based on AR(1) (left) and exchangeable (right) design correlation with $\rho=0.5$}
    \label{fig:detect_m1}
\end{figure}

The box plots of the true positive rate for different influence detection measures under different design correlation structures are given in Figure \ref{fig:detect_m1}. Under both design scenarios, we see that ACD based on SCAD and LASSO penalties yield the best results. The DF-Model is competitive but not very efficient followed closely by the original Cook's distance (CD). CW-LASSO and the local influence measure (LOC-INF) perform poorly with LOC-INF giving the worse performance. While all methods deteriorate under the exchangeable design in (b), we see that ACD remained consistent with the high performance compared to the other methods. 
\quad \\

\begin{figure}[htb!]
    \centering
    \includegraphics[width=0.49\linewidth, height=3in]{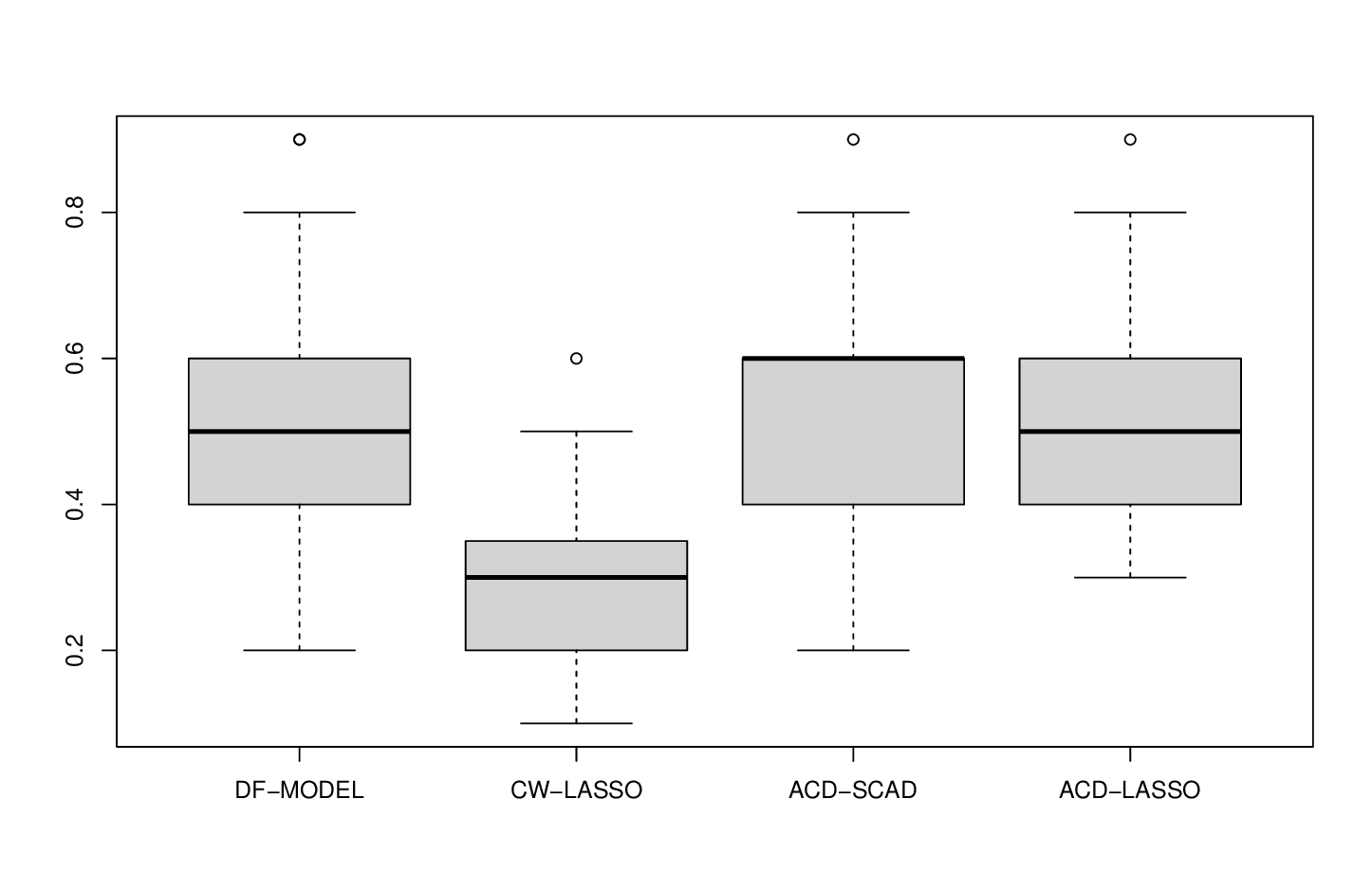}
    \includegraphics[width=0.49\linewidth, height=3in]{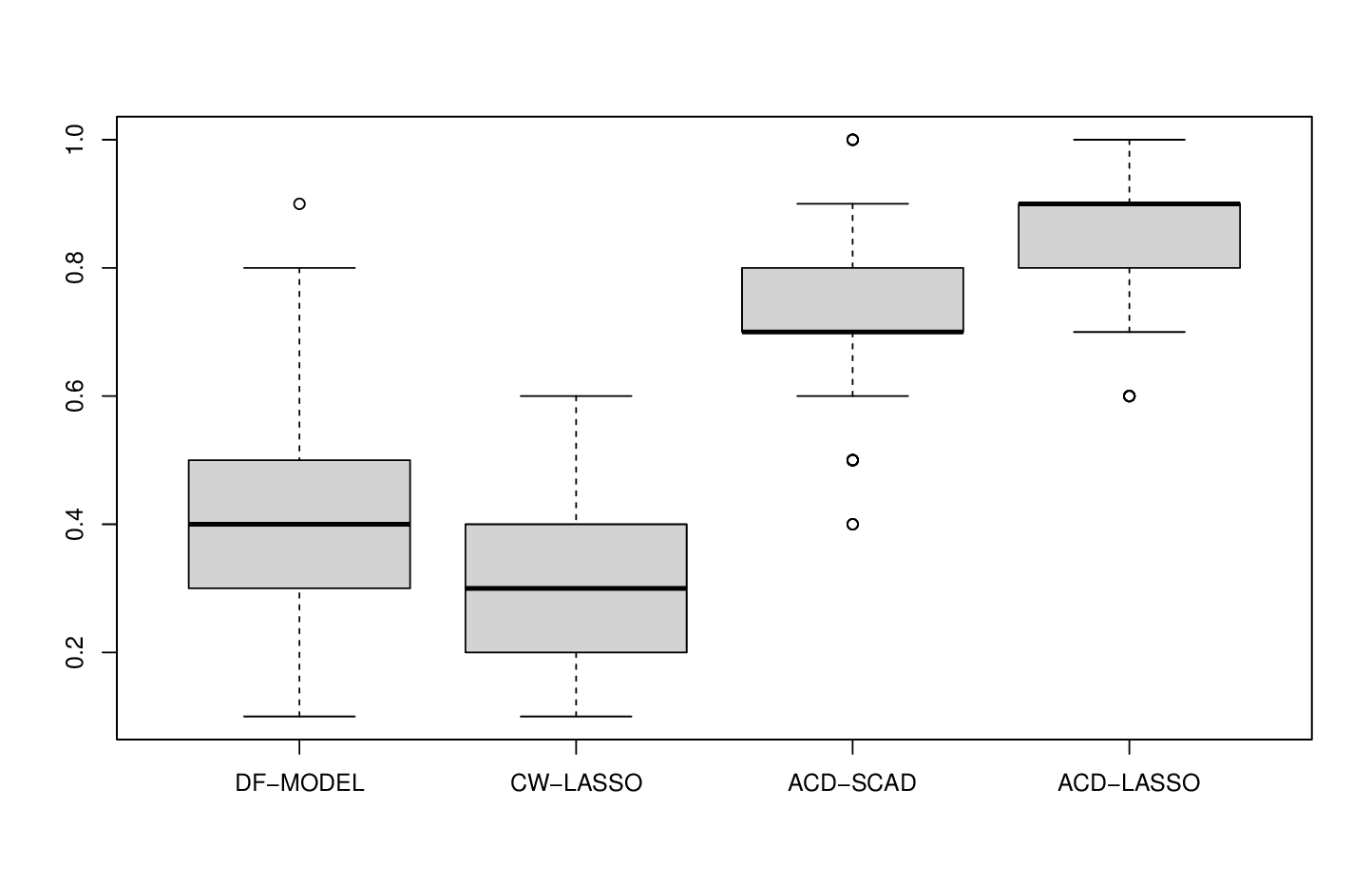}
    \caption{TPR of influence detection in high dimension based on AR(1) (left) and exchangeable (right) design correlation with $\rho=0.7$}
    \label{fig:detect_m2}
\end{figure}

{\bf Model II:} We use the same data generation process as in Model 1 but set $(n,p) = (100, 200)$, $P_{out} = 10\%$,  and $\rho = 0.7$. Also, for $\bbeta$, we allow only 5 active coefficients from the set $\{-2, -1, 0.5, 1.5, 3\}$, which are randomly assigned. Here, because $p>>n$, the classical OLS does not apply. Hence, we leave out the original Cook's distance and local influence measure. The results for 100 repeated samples are provided in Figure \ref{fig:detect_m2}.

Again, in the high dimensions we see ACD under both LASSO and SCAD penalties outperforming the DF-Model and CW-LASSO. This demonstrates the flexibility and robustness of the adaptive Cook's distance in both low- and high-dimensional settings. All examples thus far, suggest that the choice of penalty term for the sparse local estimates does not have a substantial impact on the performance of the ACD.

\subsection{Impact of Trimming Influential Observations on Variable Selection}
\begin{figure}[htb!]
    \centering
    \includegraphics[width=0.49\linewidth, height=3in]{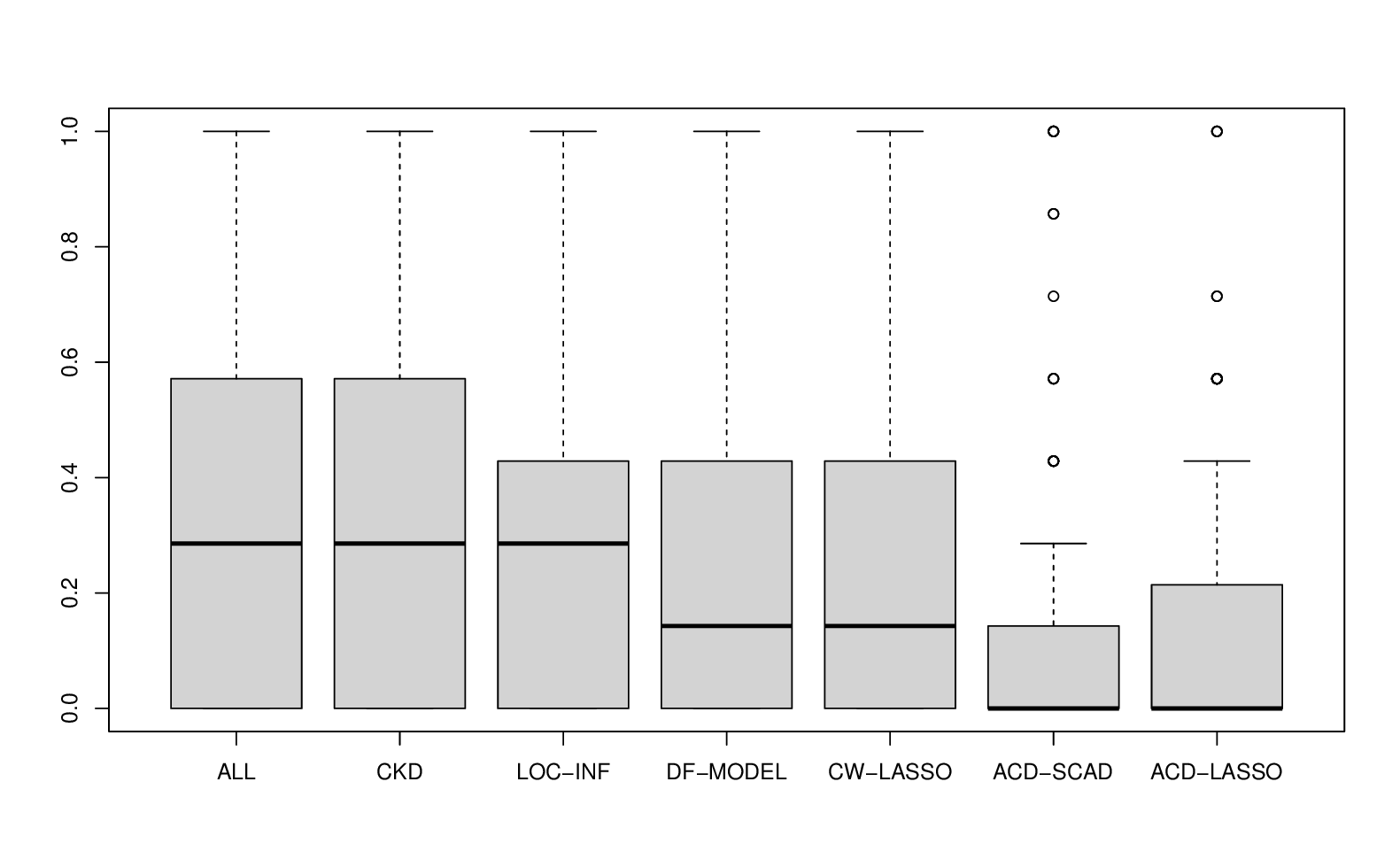}
     \includegraphics[width=0.49\linewidth, height=3in]{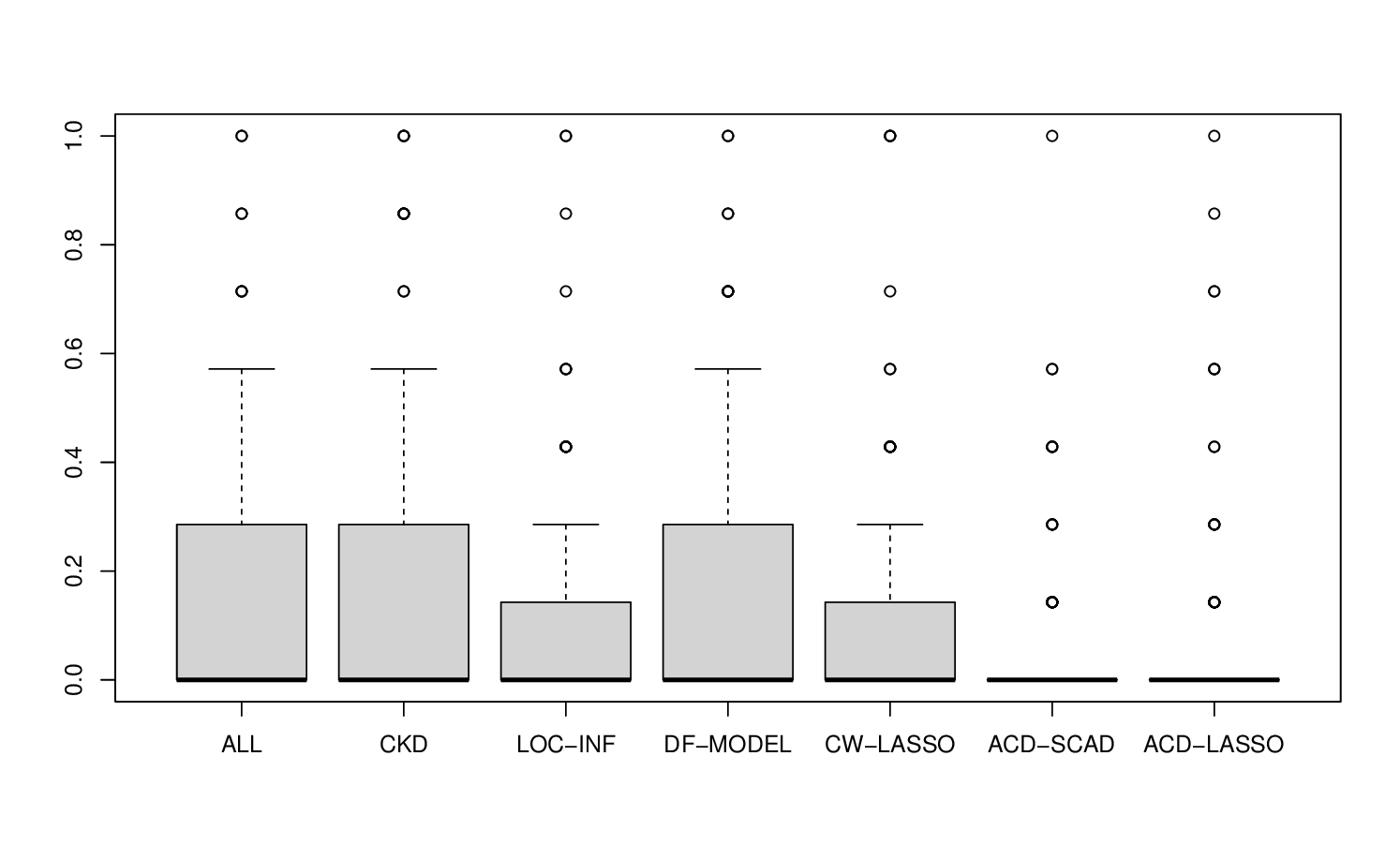}
    \caption{FPR in low dimension based on AR(1) (left) and exchangeable (right) design correlations with $\rho=0.5$}
    \label{fig:select_m1}
\end{figure}

\begin{figure}[htb!]
    \centering
    \includegraphics[width=0.49\linewidth, height=3in]{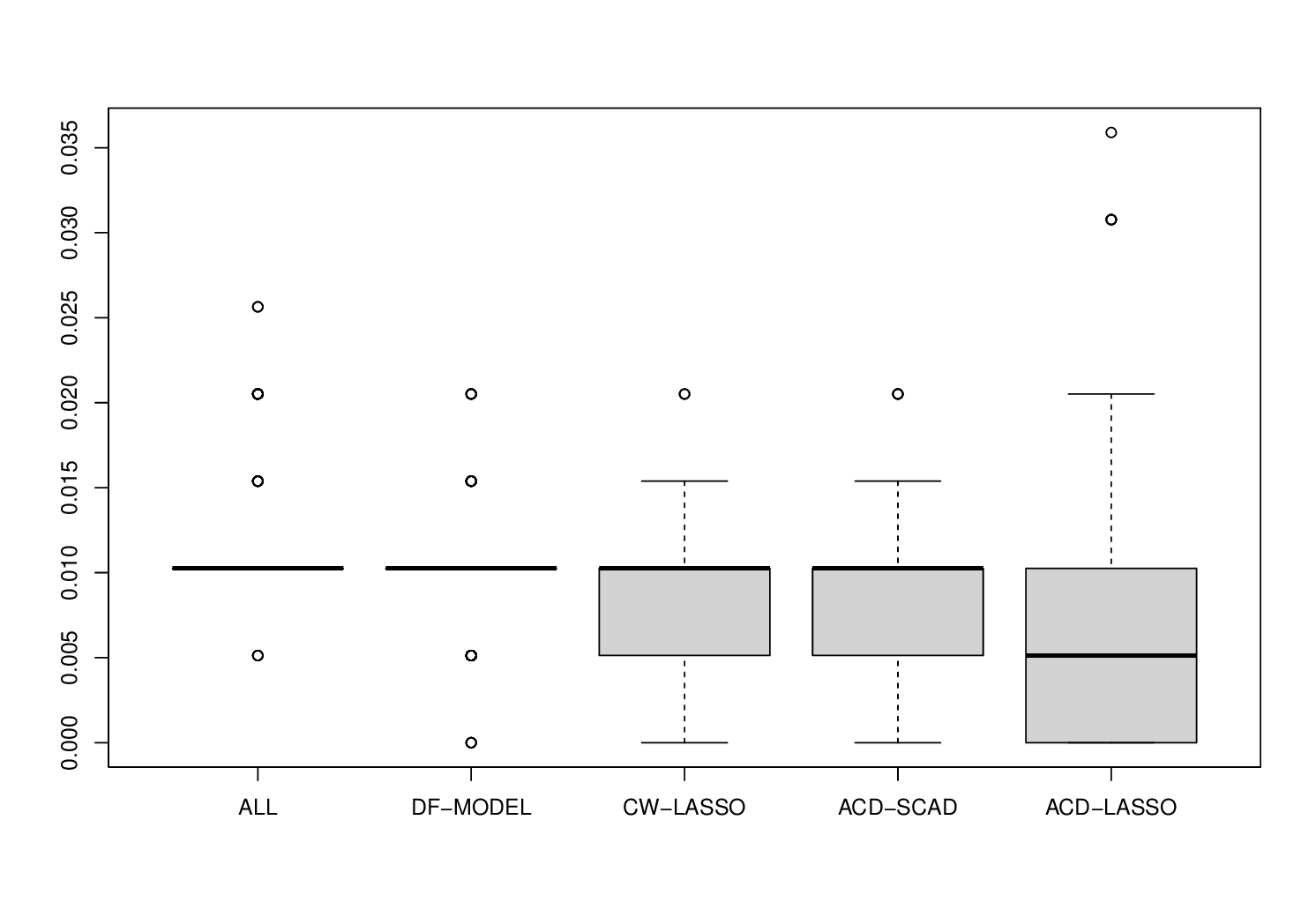}
    \includegraphics[width=0.49\linewidth, height=3in]{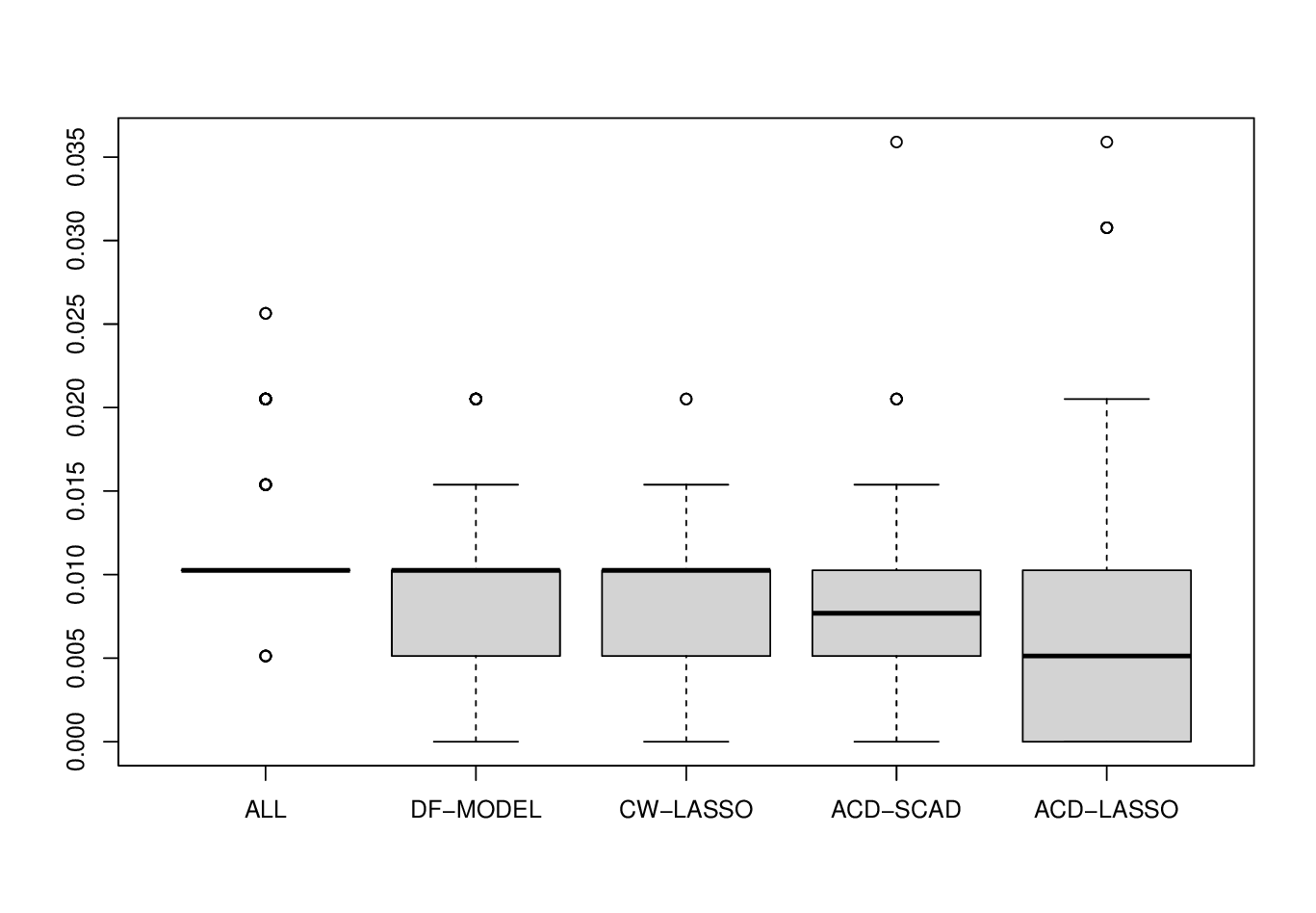}
    \caption{FPR in high dimensions based on AR(1) (left) and exchangeable (right) design correlations with $\rho=0.7$}
    \label{fig:select_m2}
\end{figure}

We study how removing observations marked as influential affects the accuracy of the variable selection. To avoid LASSO’s grouping behavior, the final selection is performed with SCAD regardless of the influence detector. The selection error is summarized by the false positive rate (FPR),
\[
\mathrm{FPR}
=\frac{\bigl|\{k\in\{1,\dots,p\}:\ \widehat{\beta}_k\neq 0 \wedge \beta_k=0\}\bigr|}
{\bigl|\{k\in\{1,\dots,p\}:\ \beta_k=0\}\bigr|},
\]
i.e., the proportion of truly null coefficients that are spuriously included. Alongside the estimator using all observations (ALL), we report results after trimming the cases identified by each detection method. Figures~\ref{fig:select_m1}--\ref{fig:select_m2} display the distribution of FPRs across Monte Carlo replicates.

In Figure~\ref{fig:select_m1}, for both AR(1) and exchangeable designs with $\rho=0.5$, the analysis that retains all observations (ALL) and the one after trimming the influential observations flagged by Cooks distance (CKD) exhibit the highest false–positive rates (elevated medians with longer upper tails). Removing cases flagged by local influence (LOC-INF) lowers the rate of spurious inclusions, and trimming guided by LASSO diagnostics (DF-MODEL and CW-LASSO) yield additional reductions. The ACD-based trimming followed by SCAD consistently delivers the smallest and most stable FPRs, with boxplots concentrated near zero. The gap between ALL and trimmed analyses is slightly larger under the AR1 design, indicating greater sensitivity to influential cases when correlations are unevenly concentrated.

Figure~\ref{fig:select_m2} considers a higher-correlation design ($\rho=0.7$). Trimming remains beneficial: every influence detector reduces FPR relative to ALL, and the ACD+SCAD combination again attains the lowest medians and the tightest dispersion. Although the absolute FPRs are small on this scale, the variance reductions are nontrivial and improve reproducibility across replicates.

Overall, we see that within the designs studied, a small number of influential observations can materially inflate false selections. Trimming flagged cases before SCAD selection generally reduces over-selection and stabilizes support recovery. The ACD-guided trimming provides the largest and most consistent FPR reductions while the model-based and coordinatewise detectors offer intermediate gains. For transparency, we recommend reporting results with and without trimming; when trimming is applied, an ACD-style screen followed by a SCAD refit gave the most stable selections in our experiments.

\section{Real Application \label{sec:5}}
\subsection{Application to Pollution Data}
We analyze the 1960 U.S. cities \emph{pollution} data, previously examined by \cite{gencc,yuzbacsi}. The sample contains $n=60$ cities with $p=15$ covariates summarizing meteorological, demographic, socioeconomic, and air-pollution features; the response is total age-adjusted mortality per $100{,}000$. Variable descriptions are provided in Table~\ref{tab:pollution_vars}. Figure~\ref{fig:corr_poll} displays the sample correlations, suggesting several highly collinear blocks.

Consistent with \cite{gencc,yuzbacsi}, we use a linear mean model. Standard residual diagnostics (residual-fitted, partial-residual trends, QQ plot) reveal no material lack of fit, so we proceed with this specification while noting that severe collinearity and a few atypical cases can still distort estimation and inference.

\begin{figure}[htb!]
    \centering
    \includegraphics[width=\linewidth,]{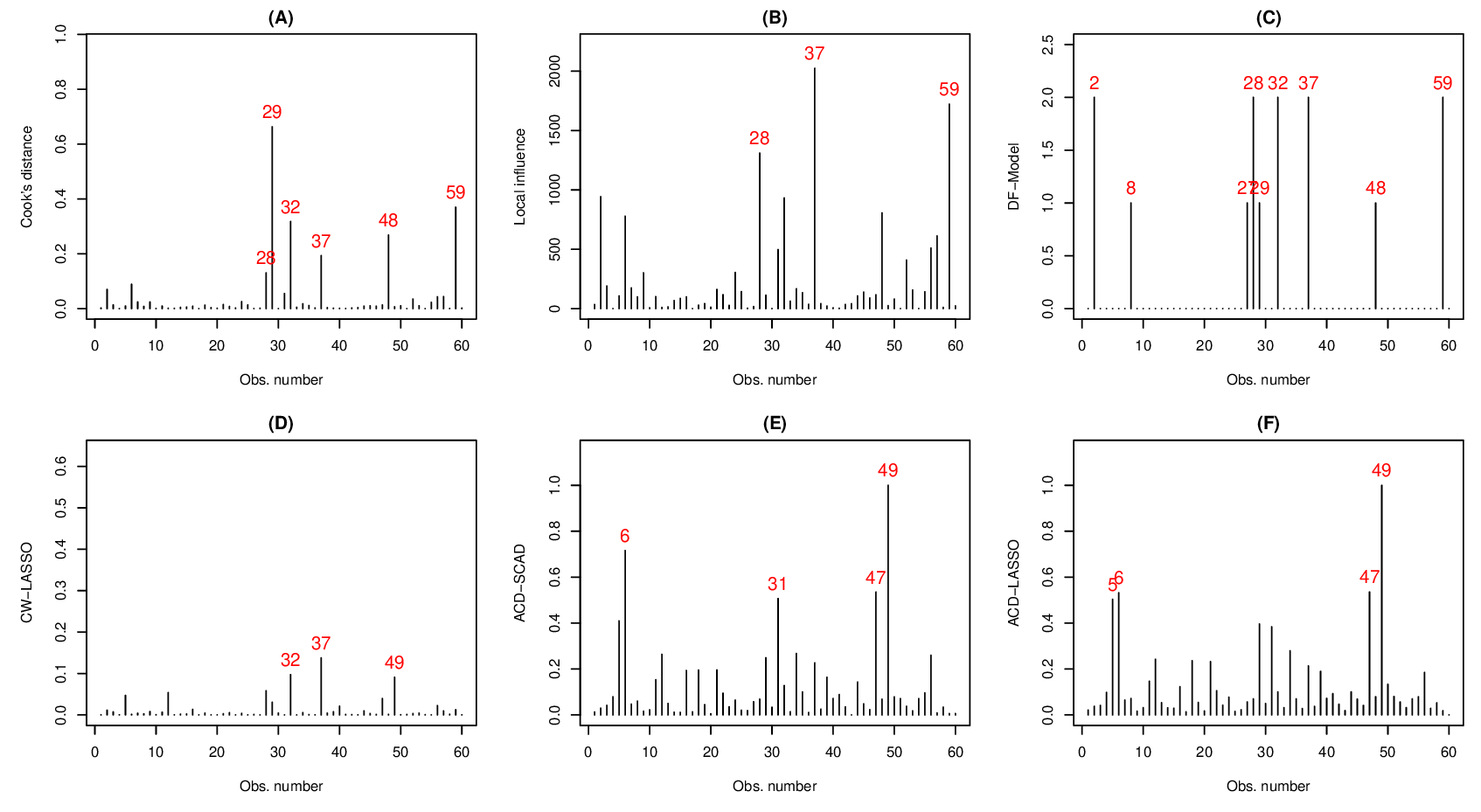}
    \caption{Influence diagnostics for pollution data}
    \label{fig:detect_poll}
\end{figure}

Following \cite{gencc}, the design exhibits extreme multicollinearity: the largest variance inflation factors are $98.64$ and $104.98$ for $X_{12}$ and $X_{13}$, respectively. Under such collinearity, conventional influence summaries must be interpreted cautiously because leverage can be large even when the combined effect on fitted values is ambiguous. In our sample, observations 18, 29, 32, 48, 49, and 59 have hat ($h_{ii}$) values that exceed $0.5$, and observations 5, 6, 12, 37, and 44 show unusually high mortality (at least $1{,}020$ per $100{,}000$). As noted by \cite{gencc}, high leverage does not by itself imply influence; a few points may dominate directionally along nearly collinear subspaces. Motivated by this, we complement classical diagnostics with our robust detection scheme and a set of competitors; see Figure~\ref{fig:detect_poll}. The procedures largely agree on a small subset of cases, but the magnitude of their estimated influence depends on how each method mitigates collinearity.

To quantify how influence and collinearity affect model composition, we conducted a stability experiment with the SCAD penalty. In each of $100$ replicates, we drew a $95\%$ subsample without replacement and computed SCAD estimates (i) on the raw subsample and (ii) after trimming the outliers flagged by each detection method. Table~\ref{tbl:sel_poll} reports selection proportions (bold when $\ge 0.50$). Several predictors—$X_{1}$, $X_{2}$, $X_{3}$, $X_{8}$, $X_{9}$, and $X_{14}$—are repeatedly chosen before and after trimming, indicating a stable core signal. By contrast, the inclusion of $X_{7}$ becomes markedly more consistent once influential cases are removed, suggesting that a small fraction of observations was suppressing its contribution under the collinear design. Methods employing $\ell_{1}$ penalties (DF-MODEL, CW-LASSO) often omit $X_{5}$ and $X_{7}$ when these are moderately correlated with $X_{6}$, behavior that is consistent with the well-documented grouping tendency of LASSO under strong predictor correlation. Trimming improves selection stability for borderline variables without altering the core set.

\begin{table}[htb!]
\caption{Proportion of selection in 100 random samples with proportions above 50\% emboldened}
\label{tbl:sel_poll}
\resizebox{\textwidth}{!}{
\begin{tabular}{cccccccc}
\hline
Predictor & ALL & CKD & LOC-INF & DF-MODEL & CW-LASSO & ACD-SCAD & ACD-LASSO   \\ \hline
$X_1$ & \textbf{0.96} & \textbf{1.00} & \textbf{1.00} & \textbf{1.00} & \textbf{1.00} & \textbf{0.91} & \textbf{0.88} \\
$X_2$ & \textbf{0.98} & \textbf{0.86} & \textbf{0.95} & \textbf{0.83} & \textbf{1.00} & \textbf{0.96} & \textbf{0.96} \\
$X_3$ & \textbf{0.91} & \textbf{0.99} & \textbf{0.99} & \textbf{0.90} & \textbf{0.85} & \textbf{0.92} & \textbf{0.86} \\
$X_4$ & 0.13 & 0.31 & 0.45 & 0.11  & 0.05 & 0.12 & 0.16 \\
$X_5$ & \textbf{0.77} & 0.34 & 0.41 & 0.19 & 0.26 & \textbf{0.64} & \textbf{0.55} \\
$X_6$ & \textbf{0.91} & \textbf{0.82} & \textbf{0.90} & \textbf{0.55} & 0.18 & \textbf{0.63} & \textbf{0.60} \\
$X_7$ & 0.19 & \textbf{0.60} & \textbf{0.84} & 0.32 & 0.11 & \textbf{0.60} & \textbf{0.63} \\
$X_8$ & \textbf{0.90} & \textbf{1.00} & \textbf{1.00} & \textbf{1.00} & \textbf{0.96} & \textbf{0.88} & \textbf{0.85} \\
$X_9$ & \textbf{1.00} & \textbf{1.00} & \textbf{1.00} & \textbf{1.00} & \textbf{1.00} & \textbf{1.00} & \textbf{1.00} \\
$X_{10}$ & 0.10 & 0.31 & 0.38 & \textbf{0.62} & \textbf{0.76} & 0.43          & 0.46 \\
$X_{11}$ & 0.00 & 0.10 & 0.34 & 0.08 & 0.09 & 0.01 & 0.02     \\
$X_{12}$ & 0.02 & 0.36 & 0.33 & 0.30 & 0.26 & 0.04 & 0.02  \\
$X_{13}$ & 0.00 & 0.31 & 0.22 & 0.10 & 0.01 & 0.01 & 0.02   \\
$X_{14}$ & \textbf{1.00} & \textbf{0.83} & \textbf{0.98} & \textbf{0.97} & \textbf{1.00} & \textbf{0.98} & \textbf{0.99} \\
$X_{15}$ & 0.08 & 0.25 & 0.46 & 0.20 & 0.09 & 0.10  & 0.15  \\ \hline
\end{tabular}
}
\end{table}

\subsection{Application to Riboflavin Data}
We evaluate all procedures—ALL, DF\mbox{-}MODEL, CW\mbox{-}LASSO, ACD\mbox{-}SCAD, and ACD\mbox{-}LASSO on a high-dimensional gene-expression dataset quantifying riboflavin (vitamin B$_2$) production in \emph{Bacillus subtilis}. The dataset was brought to prominence in the high-dimensional literature by \cite{buhlmann2014high}, is distributed via the \texttt{hdi} package in \textsf{R}, and has since been reanalyzed in several studies \citep{zhang2019pruning, gencc2021usage}. It comprises $n=71$ fermentation runs and $p=4{,}088$ predictors; each covariate is the log expression level of a gene, and the response is the log riboflavin production rate. This $p \gg n$ regime, together with pronounced gene--gene correlations, provides a stringent testbed for sparse estimation and variable selection.

\begin{table}[htb!]
\caption{Most frequently selected genes with proportions of at least 0.5 in 100 random samples are given. Proportions below 0.5 are indicated as -.}
\label{tbl:sel_pollb}
\begin{tabular}{cccccc}
\hline
Selected Gene & ALL & DF-MODEL & CW-LASSO & ACD-SCAD  & ACD-LASSO  \\ \hline
73 (ARGF)  & 0.6  & -  &  -   & -   & -   \\
{\bf 624 (LYSC)} & \bf{0.6} & -  & - & \bf{0.6} & \bf{0.6} \\
792 (PCKA) & 0.6  & -  & -  & -  & -          \\
1131 (SPOVAA) & 0.8  & - & - & - & -    \\
1762 (YEBC) & -  & -  & 0.6 & -  & -   \\
2034 (YHDZ) & 0.6  & -  & - & -  & -   \\
\bf{2564 (YOAB)} & \bf{0.8} & \bf{0.6} & - & \bf{0.6} & \bf{0.8} \\
2936 (YPZD) & - & - & 0.8 & - & -         \\
\bf{4003 (YXLD)} & \bf{0.8} & - & - & \bf{0.6} & \bf{0.6} \\
4004 (YXLE) & -  & -  & 0.8  & - & -  \\ \hline
\end{tabular}
\end{table}

\begin{figure}[htb!]
    \centering
    \includegraphics[width=\linewidth,]{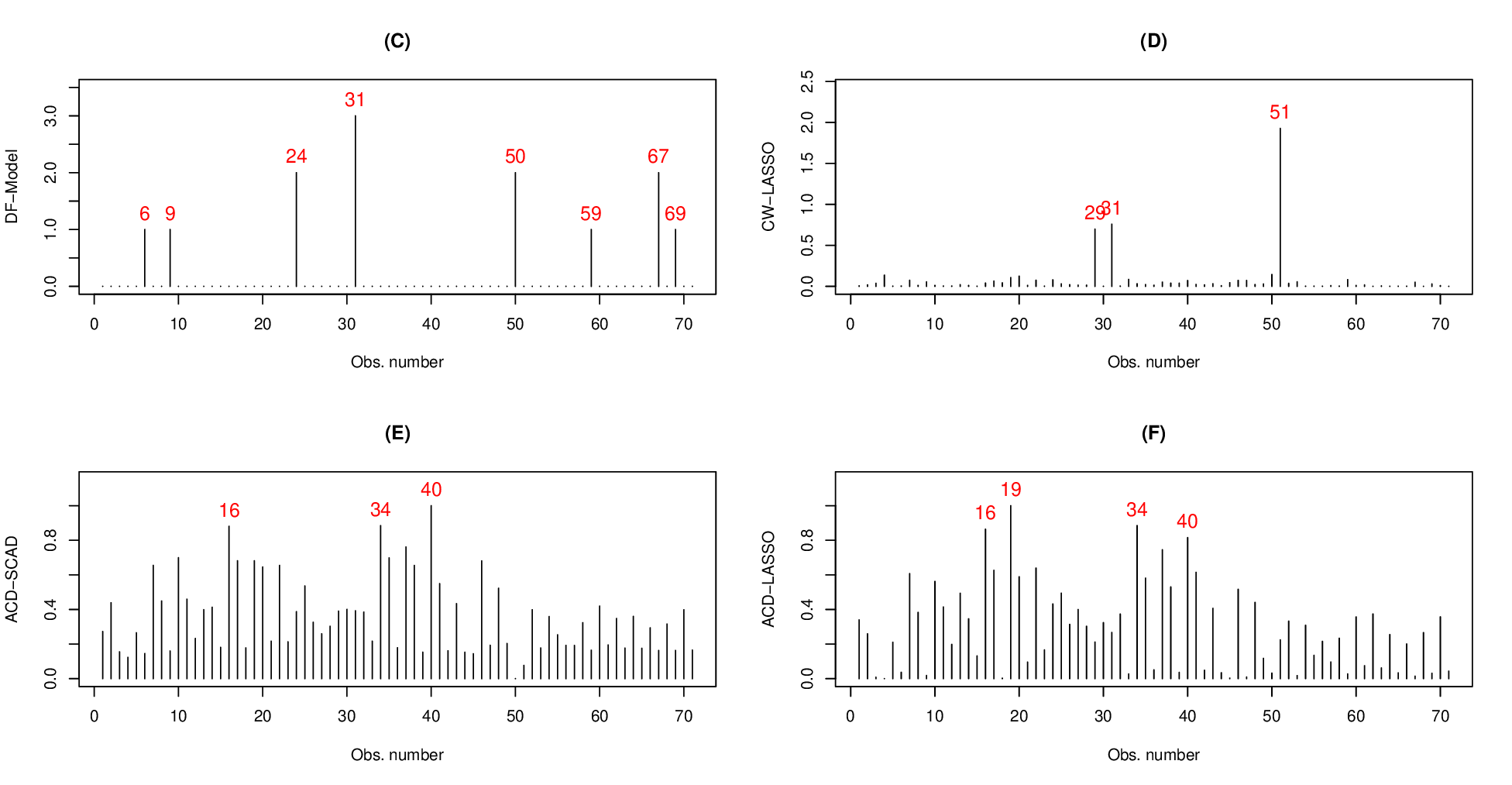}
    \caption{Influence diagnostics for Riboflavin data}
    \label{fig:detect_ribo}
\end{figure}

All four methods appear to flag different observations as influential although they all agree on some observation around 30 being influential. CW-LASSO and the ACD methods flagged fewer observations compared to DF-Model, which is not surprising based on previous results from the simulation study. We proceed with the stability of variable selection to quantify how influence and collinearity affect model composition.

Table~\ref{tbl:sel_pollb} reports, for each method, the proportion of resamples in which a gene is selected; we display only genes with a selection frequency of at least 0.5 by at least one procedure. Three signals stand out for the ACD family: \textbf{LYSC (624)}, \textbf{YOAB (2564)}, and \textbf{YXLD (4003)} are each retained by ACD--LASSO and ACD--SCAD with frequencies \(\ge 0.6\) (YOAB reaches \(0.8\) under ACD--LASSO). In contrast, the DF--Model is markedly conservative: only YOAB exceeds the 0.5 threshold, while CW--LASSO emphasizes a different subset (YEBC, YPZD, YXLE at \(0.6\)–\(0.8\)) and excludes the ACD core set. The ALL baseline, which tends to overselect, flags several genes at or above \(0.6\) (e.g., SPOVAA at \(0.8\)), but these do not replicate consistently under penalized estimators. Overall, the ACD procedures concentrate their selections on a small, repeatedly chosen group, most notably LYSC, YOAB, and YXLD, indicating greater stability under resampling relative to the CW--LASSO and DF-Model.

\section{Conclusion\label{sec:6}}
We proposed an adaptive Cook’s distance (ACD) based on sparse local gradients for influence diagnostics in high-dimensional regression models under severe collinearity and contamination. By estimating an effective index direction without specifying a link and enforcing sparsity in the local fit, ACD targets the directions along which atypical cases most perturb estimation.

In simulations that varied correlation structure and dimensionality ($p< n$ and $p> n$), ACD–LASSO and ACD–SCAD reduced masking and swamping relative to case–deletion, perturbation diagnostics, and recent $\ell_{1}$-based influence methods; trimming ACD-flagged cases yielded more stable variable selection while preserving the core signal set. Two data applications underscore these gains: in the \emph{pollution} study, trimming improved the consistency of borderline covariates while maintaining a stable core; in the riboflavin genomics study, ACD concentrated on a small, reproducible subset of genes and avoided the over/under–selection seen in baselines.

We recommend pairing ACD with resampling-based stability summaries and reporting post-trim estimates alongside raw fits. Future work includes scalable implementations for ultra-high $p$, extensions to generalized outcomes and structured penalties, and finite-sample guarantees for trimming.

\appendix
\section*{Appendix}

\begin{table}[htb!]
\centering
\caption{Variables in the 1960 U.S. Cities \emph{pollution} dataset and their index mapping. The response is indexed by $y$; predictors are $X_1,\dots,X_{15}$.}
\label{tab:pollution_vars}
\resizebox{\textwidth}{!}{
\begin{tabular}{ll p{8.5cm}}
\toprule
\textbf{Index} & \textbf{Name} & \textbf{Description (units)} \\
\midrule
$y$        & Mortality (response) & Total age-adjusted mortality per $100{,}000$ \\
\midrule
$X_{1}$  & Precipitation        & Average annual precipitation (inches) \\
$X_{2}$  & January temperature  & Average January temperature ($^{\circ}$F) \\
$X_{3}$  & July temperature     & Average July temperature ($^{\circ}$F) \\
$X_{4}$  & Relative humidity    & Annual average \% relative humidity at 1 pm \\
$X_{5}$  & Age $\ge$ 65         & \% of 1960 SMSA population aged $\ge 65$ \\
$X_{6}$  & Household size       & Average household size \\
$X_{7}$  & Education            & Median school years completed (age $>22$) \\
$X_{8}$  & Housing quality      & \% of units that are sound and fully serviced \\
$X_{9}$  & Population density   & Persons per square mile in urbanized areas (1960) \\
$X_{10}$ & Non-white            & \% non-white in urbanized areas (1960) \\
$X_{11}$ & White-collar         & \% employed in white-collar occupations \\
$X_{12}$ & HC potential         & Relative hydrocarbon pollution potential \\
$X_{13}$ & NO$_x$ potential     & Relative pollution potential of nitric oxides \\
$X_{14}$ & SO$_2$ potential     & Relative pollution potential of sulfur dioxides \\
$X_{15}$ & Low income           & \% of families with income $<\$3{,}000$ \\
\bottomrule
\end{tabular}
}
\end{table}

\begin{figure}[htb!]
    \centering
    \includegraphics[width=0.9\linewidth]{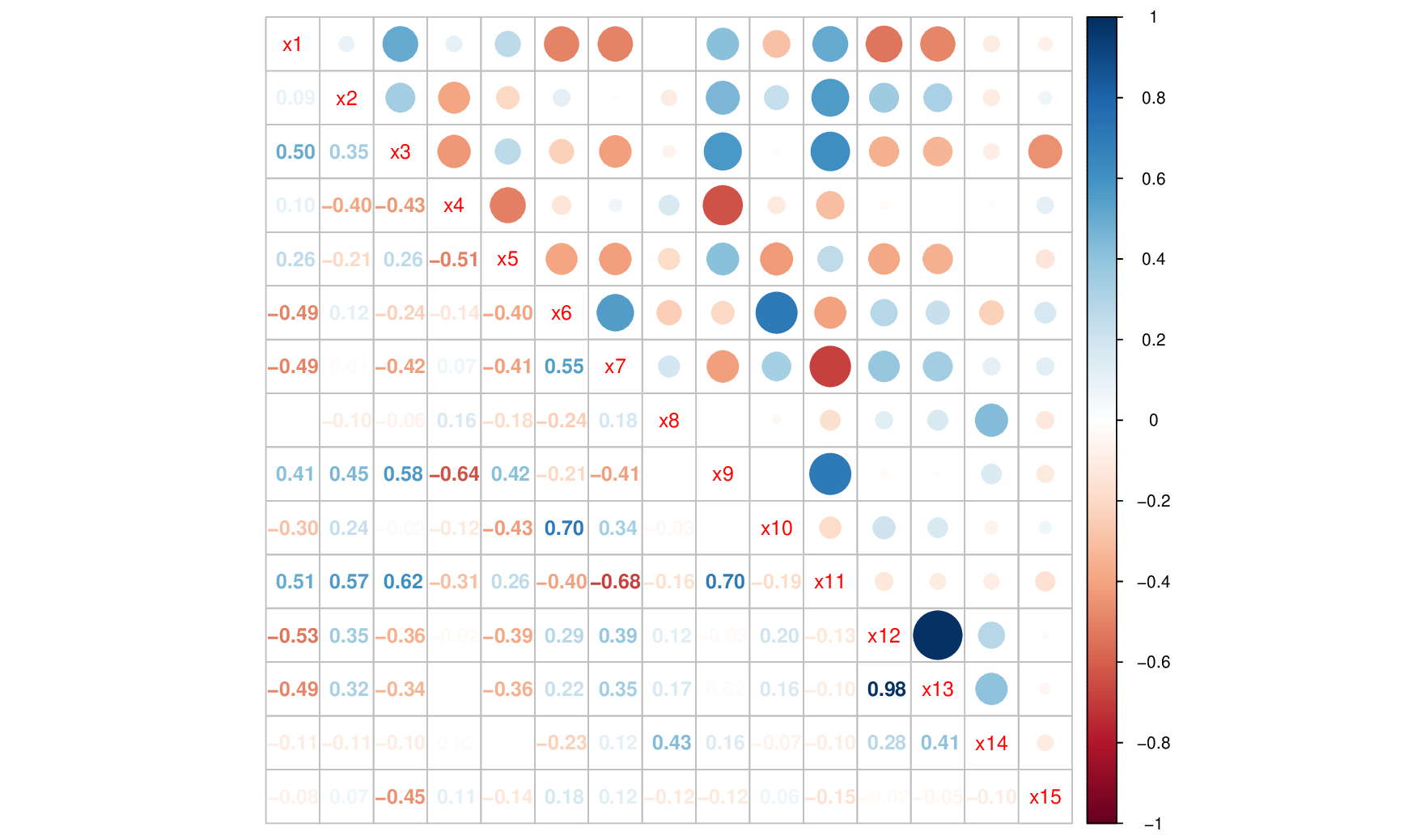}
    \caption{Design correlation matrix for pollution data}
    \label{fig:corr_poll}
\end{figure}

\quad \\

\bigskip
\begin{center}
{\large\bf Supplementary Materials}
\end{center}

\begin{description}


\item[R code:] This supplement contains the R code to reproduce the experimental studies in Section \ref{sec:4} and the real data analyses in Section \ref{sec:5}.

\end{description}





\bibliographystyle{apalike}
\bibliography{ref}

\end{document}